\documentclass[sigconf]{acmart}
\usepackage{algorithm}
\usepackage{algpseudocode}

\usepackage{multirow}   
\usepackage{booktabs}   
\usepackage{graphicx}   

\renewcommand\footnotetextcopyrightpermission[1]{}

\begin{document}

\title{HFRWKV: A High-Performance Fully On-Chip Hardware Accelerator for RWKV}


\settopmatter{authorsperrow=4}

\author{Shijie Liu}
\affiliation{%
  \institution{Sun Yat-sen University}
  \city{Guangzhou}
  \country{China}
  }

\author{Zhenghao Zeng}
\affiliation{%
  \institution{Sun Yat-sen University}
  \city{Guangzhou}
  \country{China}
  }

\author{Han Jiao}
\affiliation{%
  \institution{Sun Yat-sen University}
  \city{Guangzhou}
  \country{China}
  }

\author{Yihua Huang}
\authornote{Corresponding author. Email: huangyih@mail.sysu.edu.cn.}
\affiliation{%
  \institution{Sun Yat-sen University}
  \city{Guangzhou}
  \country{China}
  }


\begin{abstract}
RWKV is a modern RNN architecture that approaches the performance of Transformers, with the advantage of processing long contexts at a linear memory cost. However, its sequential computation pattern struggles to efficiently leverage GPU parallelism, which leads to low compute resource utilization. Furthermore, frequent off-chip weight accesses create a memory bottleneck. To address these challenges, we propose HFRWKV, an FPGA-based hardware accelerator specifically designed for RWKV. Within the matrix operation module, we propose a novel hardware-friendly hybrid-precision quantization strategy, which enhances performance while maintaining acceptable accuracy. For the complex operations including exponentiation and division, we introduce a method featuring reusable architectures combined with lookup tables or piecewise linear approximation, which is algorithmically refined to effectively balance precision and hardware resource consumption. Based on this foundation, we adopt a fully on-chip computing system integrating parallel matrix-vector processing array and an efficient pipeline architecture. Through computation reordering and chunked double buffering, it effectively eliminates data transfer bottlenecks and improves overall throughput. We implement HFRWKV on the Alveo U50 and U280 platform. Experimental results show that compared to a CPU, a throughput improvement of 63.48$\times$ and an energy efficiency improvement of 139.17$\times$. Compared to GPUs, achieves a throughput improvement of 32.33$\times$ and an energy efficiency improvement of 171.36$\times$.  
\end{abstract}

\settopmatter{printacmref=false, printccs=false}




\maketitle

\section{INTRODUCTION}
In recent years, Large Language Models (LLMs) have become a cornerstone of modern artificial intelligence, finding widespread application in diverse fields such as natural language understanding and generation, and scientific exploration \cite{LLM_infuluence}. While the predominant architecture for most state-of-the-art LLMs is the Transformer \cite{transformer, transformer1}, their self-attention mechanism’s quadratic computational and memory complexity severely limits their application in long-sequence processing and on resource-constrained devices.

To overcome the performance limitations of the Transformer architecture, both academia and industry are actively exploring novel model architectures with lower computational complexity\cite{linear_attention}. In this context, Receptance Weighted Key Value model (RWKV)\cite{rwkv-4} has emerged as a promising new algorithmic framework that ingeniously combines the parallelizable training capabilities of Transformer with the linear inference efficiency of RNN. The design allows RWKV to achieve performance comparable to mainstream Transformer models while possessing a significant efficiency advantage from its linear complexity during the inference phase. This exceptional balance of performance and efficiency has garnered widespread attention. However, the realization of RWKV at the hardware level continues to face several challenges.

(1) During inference, RWKV operates in an RNN-like mode with strong sequential dependencies across time steps, as the state computation for step t must await the result from step t-1. While GPU parallelism techniques like pipelining can overlap operations between layers, the core recurrent state update within each layer remains an unavoidable sequential bottleneck. This mismatch leads to sub-optimal resource utilization, particularly in latency-sensitive, low-batch scenarios, motivating a specialized hardware architecture.

(2) RWKV is inefficient on modern GPUs because its memory-bound element-wise operations and complex nonlinear functions, which lack dedicated hardware support, prevent it from leveraging the Tensor Cores optimized for MAC operations, resulting in significant resource underutilization.

(3) Layer Normalization\cite{layernorm} in RWKV, while stabilizing training, is a performance bottleneck because it is bandwidth-limited. Its computation of global statistics requires a multi-step reduction process on GPUs: parallel local stats are written to global DRAM, aggregated by a separate kernel, and then read back. This intensive memory traffic makes the operation’s speed dependent on memory bandwidth, not the GPU’s computational capabilities.

(4) RWKV contains multiple sequential computation modules, making its dataflow highly suitable for deeply customized pipelined architectures on FPGAs. In contrast to the fixed parallel architecture of GPUs, the reconfigurability of FPGAs allows for the design of data paths tailored to the RWKV model structure, minimizing data movement and computation latency. However, research within academia and industry on how to systematically design and implement FPGA-based hardware acceleration specifically for RWKV remains insufficient, indicating a research gap.

To address the aforementioned mismatches between the algorithm and hardware and to fully unleash the theoretical potential of the RWKV architecture, this paper proposes Hardware Friendly RWKV (HFRWKV), a fully on-chip hardware accelerator system. Ours contributions can be summarized as follows:

(1) We propose a novel, hardware-friendly hybrid-precision quantization strategy for RWKV. For the large matrix weights, we introduce an innovative $\Delta$-PoT quantization method. For all activations and additive weights, we apply a 9-bit uniform symmetric quantization.

(2) Based on the proposed $\Delta$-PoT quantization scheme, we design a reusable computing architecture that deeply integrates matrix-vector multiplication and element-wise multiplication. Centered around shifters and adders, this architecture not only efficiently executes multiplications in the $\Delta$-PoT quantized format but also supports element-wise addition.

(3) To accelerate complex functions (e.g., division, Sigmoid), we use a Piecewise Linear Approximation and Look-Up Table hybrid approach. This replaces computationally intensive operations with fast, simple linear segments and lookups, boosting speed and cutting hardware costs while preserving precision.

(4) We propose a fully on-chip architecture integrating matrix-vector, complex operation, and Layer Normalization modules. Implemented on Alveo U50/U280 platforms, its fully pipelined and highly parallel design achieves high throughput and energy efficiency. Compared to a CPU, our accelerator improves throughput by 63.48$\times$ and energy efficiency by 139.17$\times$. Against a GPU, it achieves 32.33$\times$ higher throughput and 171.36$\times$ better energy efficiency.

\section{BACKGROUND}

\subsection{RWKV Model Architecture}
\begin{figure}[h]
\captionsetup{font=small}
\centering
\includegraphics[width=65mm]{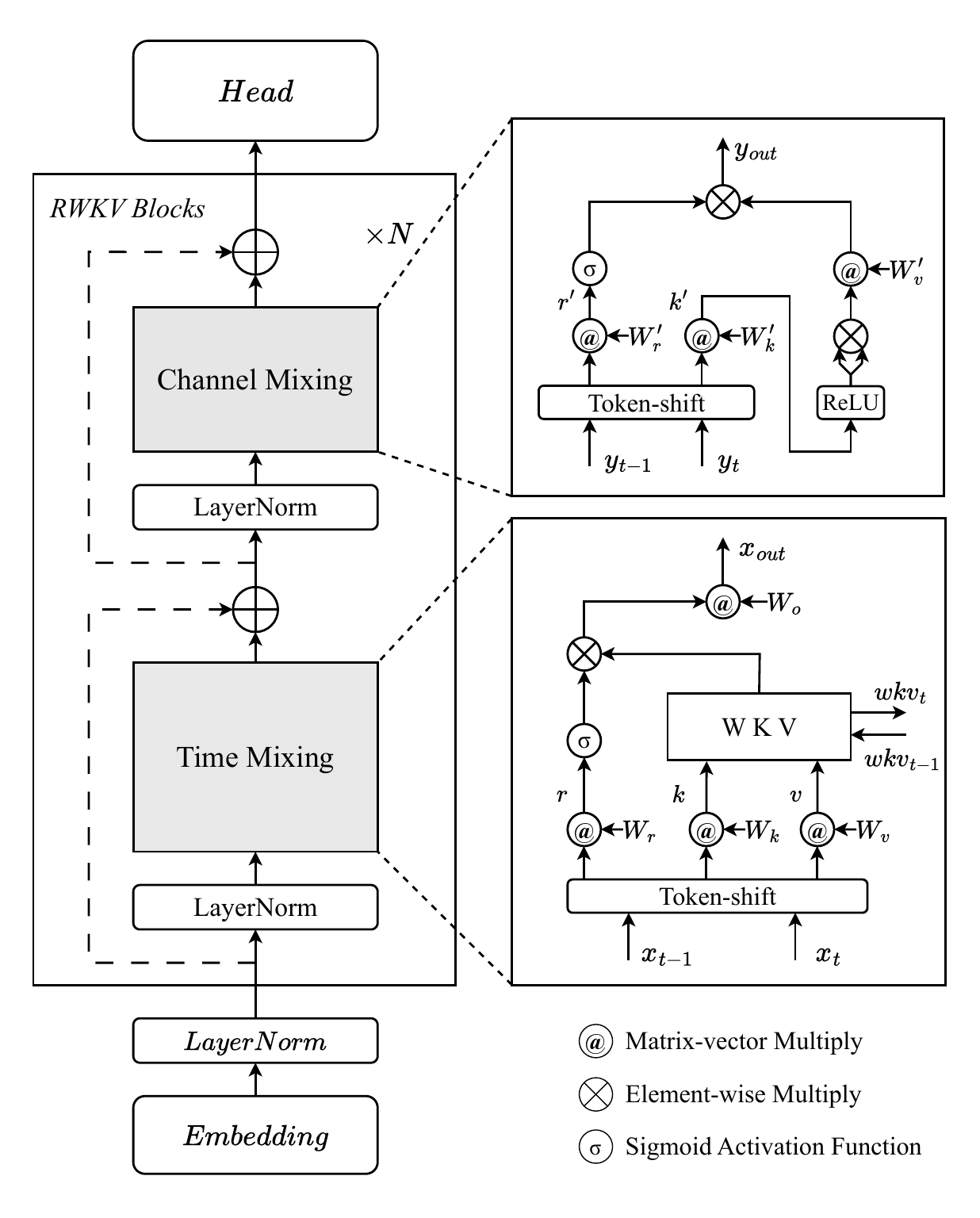}
\caption{RWKV computation process, including time mixing module and Channel mixing module.}
\label{RWKV}
\vspace{-0.2cm}
\end{figure}

RWKV combines the parallelizable training advantage of the Transformer architecture with the $O(N)$ linear inference complexity characteristic of RNNs, significantly improving efficiency in processing long sequences while maintaining high performance. As illustrated in Fig. \ref{RWKV}, the RWKV architecture is fundamentally a variant of the linear attention mechanism\cite{linear_attention1}. It consists of an initial Embedding layer, a stack of N identical residual RWKV blocks, and a final Head layer that produces logits for token prediction. Each RWKV Block is sequentially composed of a Time Mixing module and a Channel Mixing module, with each module being preceded by its own LayerNorm and a residual connection. 

Operations within each RWKV block are defined by two key computations. First, the Token-shift mechanism linearly interpolates between the current time step $x_t$ and previous time step $x_{t-1}$ to generate input vectors for the linear projections within each module. The Token-shift operation can be represented by formula (1).
\begin{equation}
    y_t^{(\lambda)} = W^{(\lambda)} \cdot (\mu^{(\lambda)} \odot x_t + (1 - \mu^{(\lambda)}) \odot x_{t-1})
\end{equation}
\noindent where $\lambda \in \{r, k, v\}$ for Time Mixing module and $\lambda \in \{r', k'\}$ for Channel Mixing module, and $\odot$ denotes element-wise multiplication.

Second, the WKV computation in the Time Mixing module acts as a weighted average, recurrently updating its state $wkv_t$ based on the current key $k_t$, value $v_t$, a trainable time decay vector $w$, and a bonus vector $u$ for the current token, which can be represented by formula (2).
\begin{equation}
    wkv_t = \frac{\sum_{i=1}^{t-1} e^{-(t-1-i)w+k_i} \odot v_i + e^{u+k_t} \odot v_t}{\sum_{i=1}^{t-1} e^{-(t-1-i)w+k_i} + e^{u+k_t}}
\end{equation}

\subsection{Quantization Schemes for LLMs}

In the research of LLMs, Model Quantization is an effective technique for improving deployment and inference efficiency. Uniform Quantization \cite{uniform_quantization} is the most fundamental quantization method, which linearly maps a floating-point number $x$ to an integer $q$ with $b$ bits. Uniform Quantization imposes a fixed representational allocation, assigning equal precision across dense and sparse regions of the distribution, which amplifies outlier-induced errors under low-bit settings and leads to substantial accuracy degradation. 

To address the drawbacks of uniform quantization, researchers have explored non-uniform quantization methods \cite{non_uniform_quantization}\cite{non_uniform_quantization_2}\cite{non_uniform_quantization_3}, such as logarithmic quantization\cite{Log_Quantization}\cite{Log_Quantization_2}. However, general non-uniform quantization schemes face challenges in hardware implementation. To address the drawbacks, researchers have explored non-uniform quantization methods that match the distribution of quantization levels with the statistical distribution
of the weights. At the hardware level, the multiplication of a floating-point activation with an arbitrarily non-uniformly quantized weight typically requires complex logic circuits, which can offset some of the performance advantages gained from quantization \cite{MIX_Q_inHardware_difficulty}.

To strike a balance between non-uniformity and hardware efficiency, Powers-of-Two (PoT) \cite{PoT}  quantization was proposed. PoT maps a continuous floating-point weight tensor $w$ to a discrete set composed of powers-of-two values. The mathematical formulation is shown in formula (3):
\begin{equation}
    w_q = S \cdot \operatorname{sign}(w) \cdot 2^E
\end{equation}

\noindent where $w_q$ is the quantized weight, $S$ is the floating-point scale factor, $\operatorname{sign}(w)$ denotes the sign of the original weight $w$, and $E$ is an integer representing the quantized exponent, whose value range is determined by the quantization bit-width $b$.

Although PoT quantization is highly efficient, its quantization levels are limited to a single power-of-two term, resulting in limited representational capacity. To address this, Li et al. proposed Additive Powers-of-Two Quantization(APoT) \cite{APoT}. Following the introduction of APoT, subsequent research, such as the "n-hot" quantization \cite{nhot_PoT}, which introduces bit-level sparsity and incorporates both additive and subtractive PoT terms to improve representational capacity, has extended the concept of additive quantization to better balance the trade-off between hardware efficiency and model precision. These advancements demonstrate a clear trend towards developing more flexible, hardware-aware non-uniform quantization methods.

\section{QUANTIZATION STRATEGY}
To reduce computational complexity and memory overhead, and improve hardware efficiency, this paper proposes a hardware-friendly mixed-precision quantization architecture. For weights involved in multiplication operations, a novel $\Delta$-PoT quantization is adopted to alleviate the bottleneck caused by multiplication. In contrast, for weights used in element-wise operations such as addition, a 9-bit uniform symmetric quantization is employed to ensure precision. Furthermore, 9-bit uniform fixed point quantization is applied to all other activations and intermediate results to achieve the optimal balance between performance and accuracy.

\subsection{Quantization Scheme for Matrix-Vector Multiplication}
Observing that matrix-vector multiplications constitute a major computational bottleneck in the RWKV model and impose significant memory bandwidth pressure due to the matrix weights, this work proposes a novel quantization scheme for multiplication and an algorithm-aware architecture tailored for the matrix-vector product. Inspired by PoT quantization, we introduce an improved variant, $\Delta$-PoT, which enhances the APoT scheme by employing a differential encoding format for weight storage.


In APoT quantization, we only consider unsigned numbers for simplicity: each level is the sum of $n$ PoT terms as shown below.
\begin{gather}
Q^a (a, kn) = \gamma \times \left\{ \sum_{i=0}^{n-1} p_i \right\} 
\\ \nonumber
\text{where} \quad
p_i \in \left\{ 0, \frac{1}{2^i}, \frac{1}{2^{i+n}}, \dots, \frac{1}{2^{i+(2^k-2)n}} \right\}
\end{gather}

Here, $\gamma$ is a scaling coefficient to make sure the maximum level in $Q^a$ is $a$. $k$ denotes the bit-width for each additive term, and $n$ represents the number of additive terms. Given total bit-width $b$ and base bit-width $k$, $n$ is calculated as $n = b/k$, yielding $2^{kn} = 2^b$ total levels.

Different from APoT, our $\Delta$-PoT quantization allows each $p_i$ to have distinct $k_i$, with quantization levels defined by:
\begin{gather}
Q^a \left(a, \sum k_i \right) = 2\gamma \times \left\{ \sum_{i=0}^{n-1} p_i \right\} 
\\ \nonumber \text{where} \quad 
p_i \in \left\{ 0, \frac{p_{i-1}}{2^1}, \frac{p_{i-1}}{2^2}, \ \dots, \frac{p_{i-1}}{2^{2^{k_i}-1}} \right\},p_{-1}=1
\end{gather}
This approach permits arbitrary allocation of $k_i$ values rather than being constrained by $k = b/n$, providing significant advantages for flexible bit-width selection on FPGAs and enabling higher precision in certain scenarios.

Setting $p_i = 1/2^{q_i}$, $\Delta q_i = q_i-q_{i-1}$, we derive:
\begin{equation}
\begin{aligned}
p_i = 
\begin{cases} 
p_{i-1} \cdot 2^{-\Delta q_i}, &\Delta q_i > 0, \\
0,    &\Delta q_i = 0 ,
\end{cases}
\quad 
\Delta q_i \in \{ 0, 1, 2, \dots, 2^{k_i}-1 \}
\end{aligned}
\end{equation}
Here, $\Delta q_i$ represents the difference-encoded value stored for each term, so we call it $\Delta$-PoT quantization.

We use $b = 4$ and $k = 2$ as an example to analyze the advantages of $\Delta$-PoT relative to APoT. For this example, APoT have $p_0 \in \{0, 2^0, 2^{-2}, 2^{-4}\}$, $p_1 \in \{0, 2^{-1}, 2^{-3}, 2^{-5}\}$, and $\Delta$-PoT have $p_0 \in \{0, 2^{-1}, 2^{-2}, 2^{-3}\}$, $p_1 \in \{0, 2^{-1}p_0, 2^{-2}p_0, 2^{-3}p_0\}$. If we need to quantify the value of $\gamma \times (2^0 + 2^{-2})$, APoT only can be $\gamma \times (2^0 + 2^{-3})$, but $\Delta$-PoT can precisely represent $2\gamma \times (2^{-1} + 2^{-3})$.

The $\Delta$-PoT quantization scheme we propose eliminates the dependency on multipliers. It can replace DSP units through a combination of several shifters and adders, which offers great convenience for hardware platforms that support barrel shifting, such as FPGAs. First, it reduces computational complexity. Second, it replaces complex multipliers with a simple “shift-add” structure, saving on-chip resources. Third, it performs computation in a form that best aligns with hardware characteristics, reducing computational latency while ensuring model performance through high-precision, non-uniform representation. Finally, $\Delta$-PoT allows weights to be stored with a low bit width, saving storage space and memory bandwidth, while its differential encoding mechanism also covers a larger numerical representation range.

Compared to uniform fixed-point quantization, different weights can carry different scaling factors, further increasing bandwidth pressure and complicating operators. In contrast, our quantization scheme does not require any modification of the operators and is hardware-friendly.
\subsection{Mixed-Precision Quantization for Element-wise Operations}
Except for the matrix weights that are quantized using $\Delta$ -PoT, weights that require multiplication with activation values also adopt $\Delta$-PoT quantization. Apart from this, all weights involved in addition operations with activation values utilize 9-bit uniform symmetric quantization. This is because addition operations cannot easily leverage the structural advantages of $\Delta$-PoT and the weights in the RWKV model are relatively more concentrated in value. Therefore, considering the accuracy requirements, a simple and direct uniform quantization is a suitable choice.

To ensure precision against outliers, we apply 9-bit uniform quantization to all activations. For high-precision operations like division and Sigmoid, 9-bit fixed-point quantization is sufficient as they are weightless. To further guarantee numerical stability and mitigate errors, their hardware modules operate internally at 16-bit precision.

\section{ARCHITECTURE DESIGN}
\subsection{RWKV Hardware Architecture}

\begin{figure}[htbp] 
\captionsetup{font=small}
\centering
\includegraphics[width=85mm]{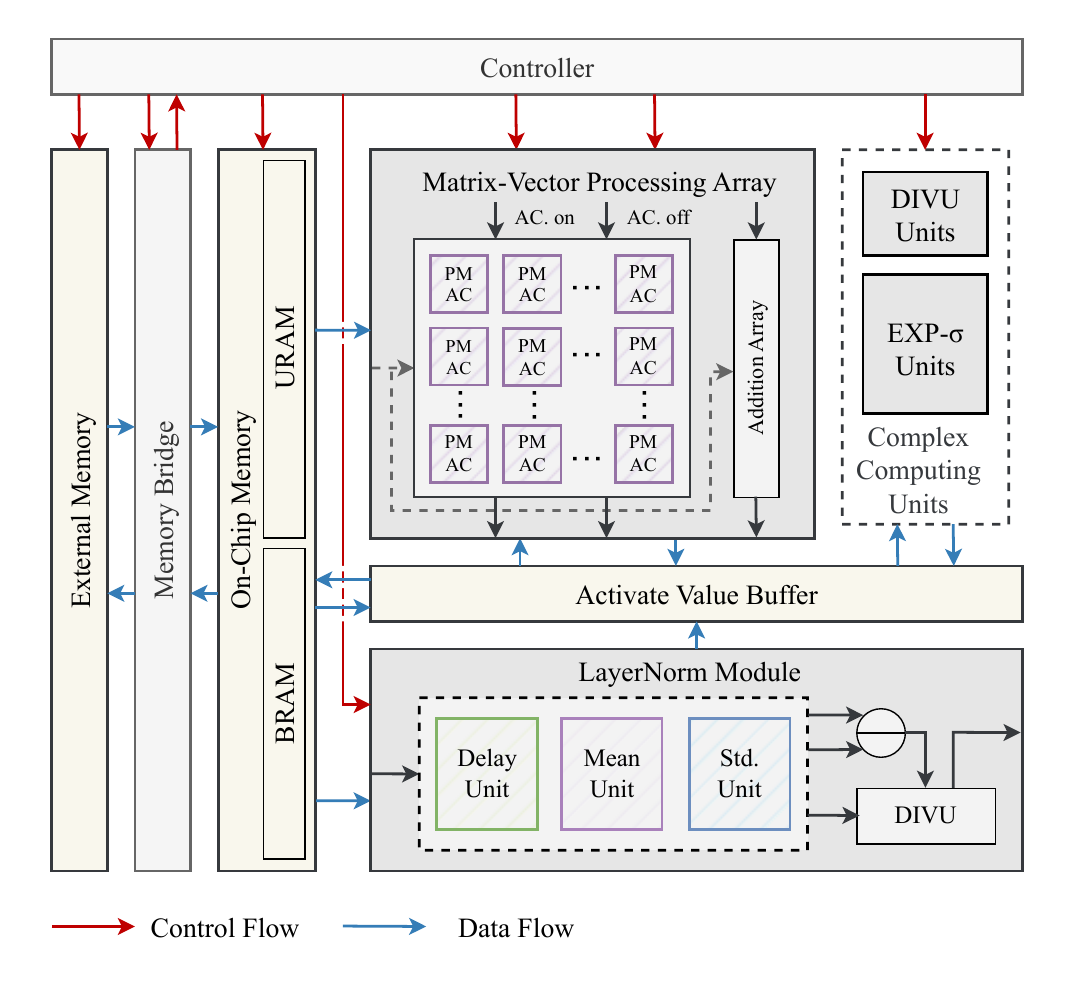}
\caption{HFRWKV System Architecture}
\label{System Architecture}
\vspace{-0.5cm}
\end{figure}

To accelerate the entire RWKV architecture, specialized hardware support is required for complex operations such as division, exponentiation, sigmoid activation, and LayerNorm. Division and exponentiation are primarily concentrated in the WKV operator, while sigmoid is involved in every \textit{v}-term computation. LayerNorm operations must be performed before inputs enter the Time Mixing and Channel Mixing modules. To address these requirements, we designed a fully on-chip computing system tailored for the RWKV model, as illustrated in Fig. \ref{System Architecture}.

The architecture comprises an External Memory, Memory Bridge, On-Chip Memory, Controller, Activate Value Buffer, Matrix-Vector Processing Array, Complex Computing Units, and a LayerNorm Module. Off-chip weights are transferred in bulk from external memory to on-chip memory via the Memory Bridge, ensuring efficient data access for computation. The Controller orchestrates data transfers from external memory and manages on-chip modules to fetch data from designated addresses in on-chip memory.

To minimize off-chip memory access, vector weights and historical values are stored entirely in on-chip BRAMs. Since weights are quantized with mixed precision, they are concatenated off-chip and decoded to the corresponding bit-width after being transferred on-chip. However, large matrix weights (e.g., 4096$\times$4096 elements for the model size of 7B) exceed on-chip capacity. We therefore employ a ping-pong double-buffering strategy using URAMs, enabling chunked matrix transfers that overlap with computation. This allows full storage for smaller models (e.g., 0.43B) in URAMs and block-wise transfers for larger ones (e.g., 7B), effectively hiding memory latency and fully utilizing HBM bandwidth.

According to the structure of the RWKV model, LayerNorm must be applied to input vectors before major computation blocks. Therefore, the data flow is directed from the on-chip memory to the LayerNorm Module. Based on the computational logic of layer normalisation, this module incorporates a Delay Unit for pipeline alignment, a Mean Unit, and a Standard Deviation (Std) Unit, followed by subtraction and division operations ($\text{DIVU}$) to generate normalized outputs.

\begin{figure}[b] 
\captionsetup{font=small}
\centering
\includegraphics[width=85mm]{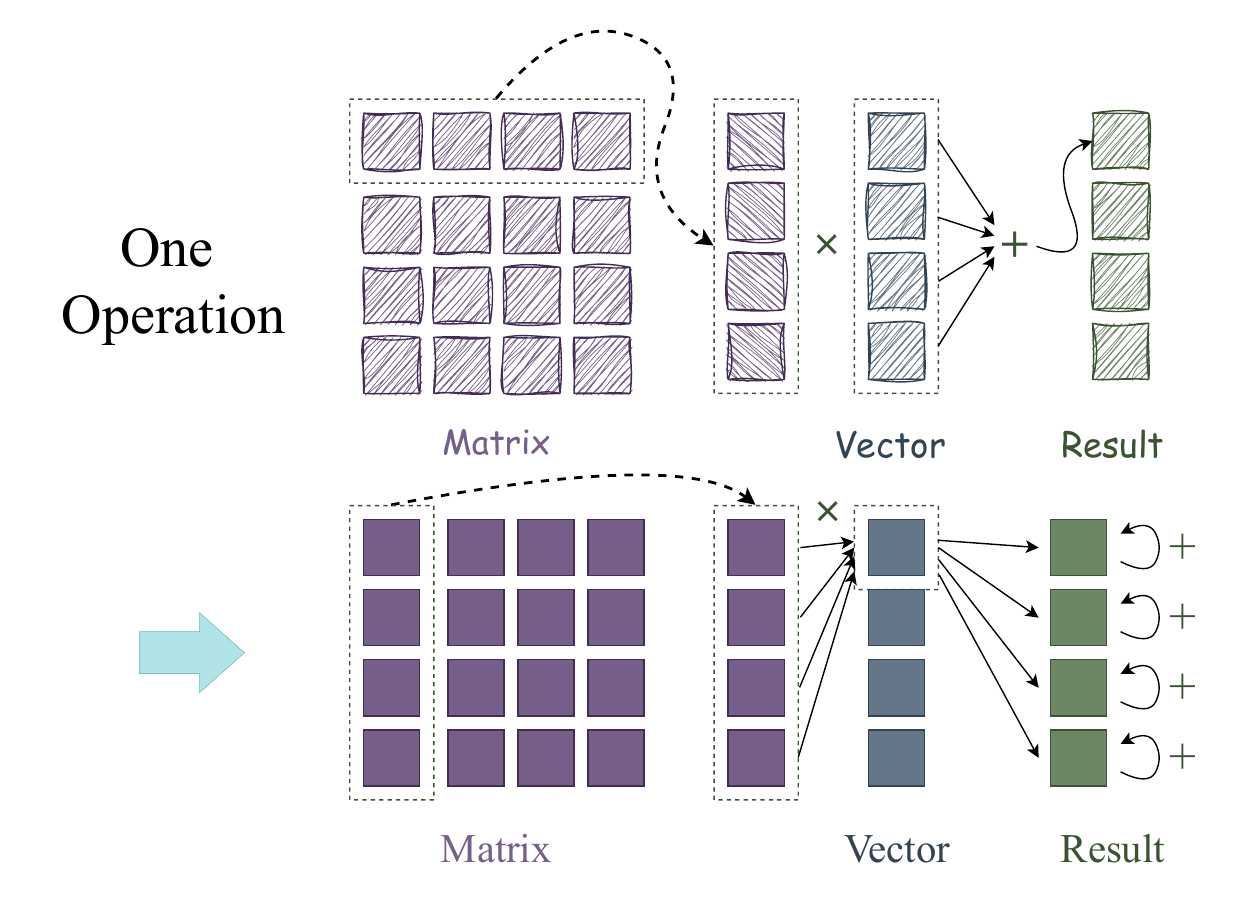}
\caption{Matrix-Vector Multiply Conputation}
\label{Matrix Multiply}
\end{figure}

Normalized vectors are stored in the Activate Value Buffer for subsequent computations. This buffer temporarily retains data with temporal/spatial locality, including activations, intermediate variables, and prefetched weights.

Matrix-vector multiplication and element-wise operations are accelerated by the Matrix-Vector Processing Array. This reusable structure supports batched parallel computation with three operational modes: Matrix-vector multiplication enabled when accumulators (AC) in PMAC units are activated; Element-wise multiplication activated when AC units are disabled; Addition selected via the adder array.

Complex operations in the WKV operator (e.g., exponentiation and division) are handled by dedicated Complex Computing Units, which include an Unsigned Division Unit and a shared Natural Exponent-Sigmoid Unit. These modules implement all non-linear computations in RWKV with resource-optimized designs.

All computational modules feature fully pipelined, highly parallel architectures. Fine-grained pipelining enables batched processing of element-wise operations, while coarse-grained pipelining overlaps data transfer with computation.

\begin{figure*}[t] 
\captionsetup{font=small}
\centering
\includegraphics[width=\textwidth]{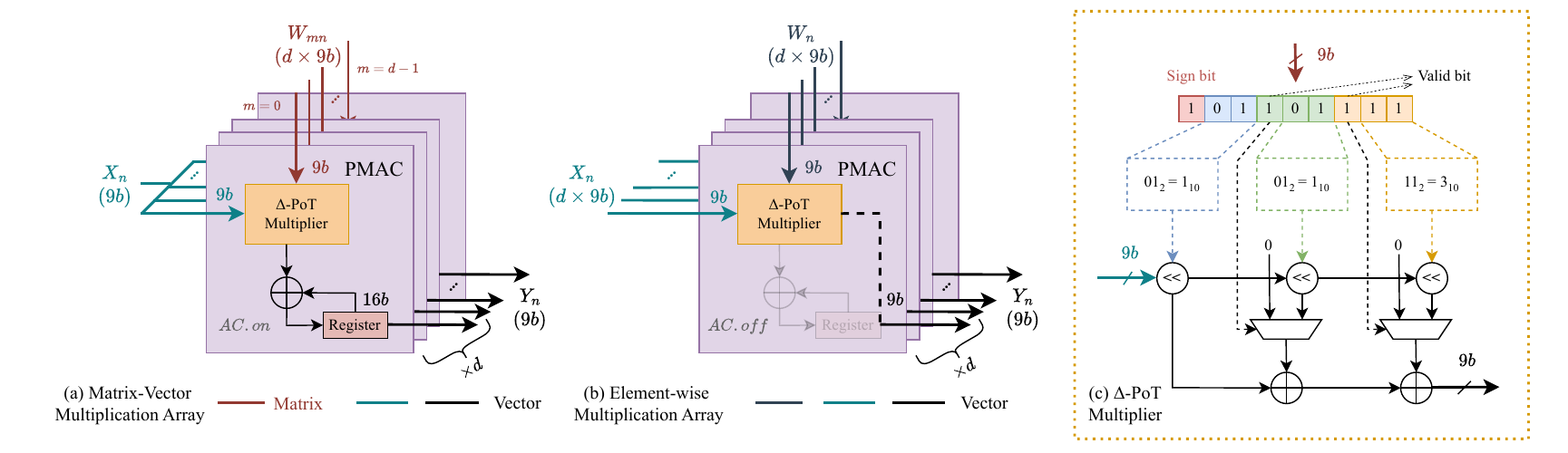}
\caption{Matrix-Vector Processing Array}
\label{GEMV}
\vspace{-0.2cm}
\end{figure*}

\subsection{Matrix-Vector Processing Array}

We propose specialized architectural optimizations for matrix-vector multiplication (MVM) computations using the $\Delta$-PoT representation, introducing a parallel matrix-vector processing array. Inspired by Systolic Array\cite{Systolic_Array}, we observe that MVM computations can similarly exploit temporal and spatial locality. Crucially, temporal locality can be \textit{maximally exploited} through extreme parallelism: all computations involving current data elements are completed synchronously within a single cycle, achieving single-fetch data reuse. Spatial locality patterns are illustrated in Fig.~\ref{Matrix Multiply}.

Consider a square matrix $W \in \mathbb{R}^{n\times n}$ and vector $\mathbf{v} \in \mathbb{R}^{n}$. As shown in Fig.~\ref{Matrix Multiply}, the first element of the result vector $\mathbf{r}(1) = \sum_{j=1}^{n} W_{1j}v_j$. Decomposing this computation reveals that $v_1$ must multiply every element in $W$'s first column ($W_{i1} \forall i$). By reordering operations, we process all $v_1 \times W_{i1}$ computations concurrently within one cycle. This approach achieves High parallelism of $O(n)$ operations per cycle and Minimal data access of single fetch per vector element. The same principle extends to non-square matrices through dimension-aware scheduling.

The proposed architecture employs a parallel matrix-vector processing array utilizing $\Delta$-PoT Multiplication Accumulators (PMAC) as computational units, implemented with a three-stage pipeline. Fig.~\ref{GEMV}(a) shows the matrix-vector multiplication mode with activated accumulators (AC), where each clock cycle processes column $n$ of the matrix and element $n$ of the vector. To prevent overflow during accumulation, 16-bit registers are incorporated in the accumulators. For $d$ parallel PMAC units processing $l$-dimensional inputs, the result latency is $(l+4)\times(l/d)$ cycles, accounting for pipeline initialization and drainage overhead. Fig.~\ref{GEMV}(b) demonstrates the element-wise multiplication mode with deactivated AC, processing vector element $n$ per cycle. This configuration computes element-wise products with $(l/d+4)$ cycle latency, achieving full pipeline utilization. Fig.~\ref{GEMV}(c) Shows the structure of the  $\Delta$-PoT Multiplier. It decomposes multiplier inputs (excluding sign bits) into three components processed through barrel shifters and shift-add accumulation. This design replaces conventional DSP-based multipliers with optimized combinational logic, significantly reducing DSP resource utilization while maintaining numerical precision. All computational paths incorporate overflow protection mechanisms not explicitly shown in the diagram.

\begin{algorithm}[b]
\small 
\caption{Leading One Detector using hierarchical binary search algorithm}
\label{alg:lod}
\begin{algorithmic}[1]
\Function{LOD}{$in[k-1:0]$}
  \State $p \gets 0,\ w \gets k,\ d \gets in$
  \While{$w > 1$}
    \State $h \gets w/2$
    \If{$\bigvee d[w-1:h] = 1$}
      \State $d \gets d[w-1:h],\ p \gets p + h$
    \Else
      \State $d \gets d[h-1:0]$
    \EndIf
    \State $w \gets h$
  \EndWhile
  \If{$d = 1$}
    \State \Return $p$
  \Else
    \State \Return $-1$
  \EndIf
\EndFunction
\end{algorithmic}
\end{algorithm}

\subsection{Unsigned Division Unit}

\begin{figure}[b]
  \centering
  \includegraphics[width=90mm]{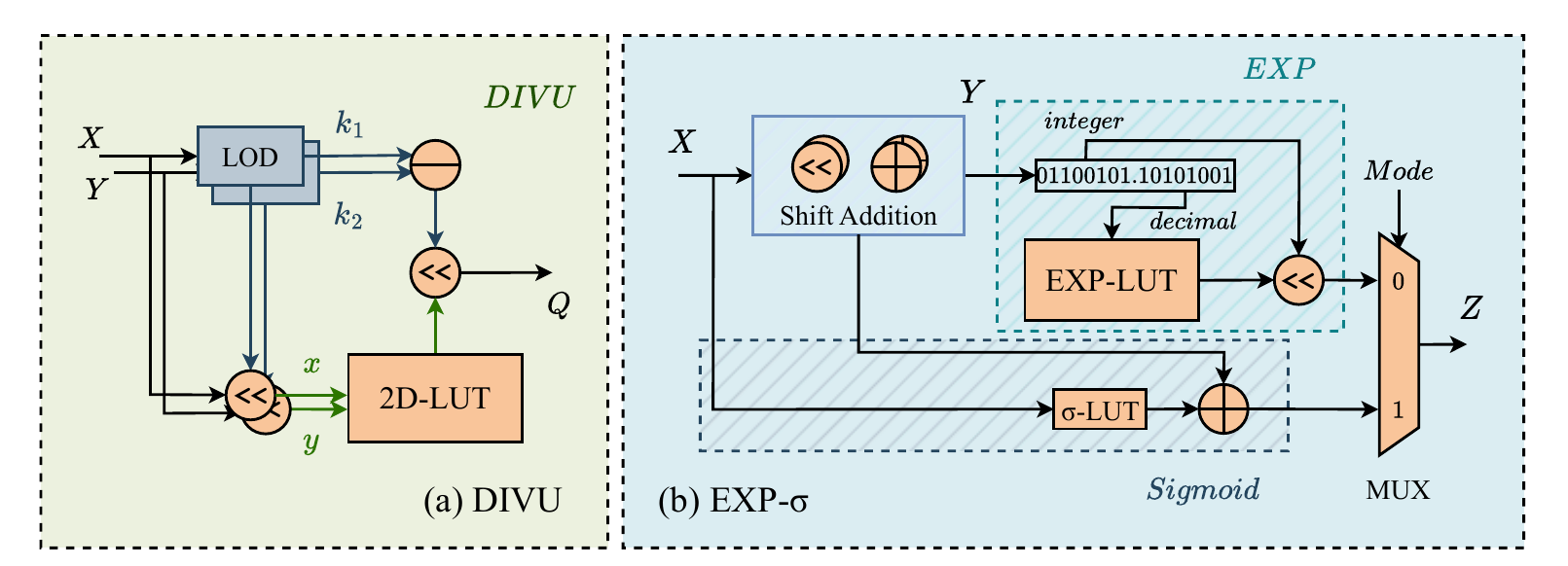}
  \centering
  \caption{(a)Unsigned Division Unit (b)Exponential–Sigmoid Unit}
  \label{DIVU_EXP_Sigmoid}
  \vspace{-0.5cm}
\end{figure}

The unsigned division unit (DIVU) in our design, illustrated in Fig.~\ref{DIVU_EXP_Sigmoid}(a), requires sign-bit separation and signed-to-unsigned conversion prior to operation. The division algorithm decomposes dividend $X$ and divisor $Y$ into normalized forms $X = 2^{k_1} \cdot x$ and $Y = 2^{k_2} \cdot y$, where $1 \leq x,y < 2$. The quotient is then derived as:
\begin{equation}
Q = \frac{X}{Y}= \frac{x}{y} \times 2^{(k_1 - k_2)} = \frac{x}{y} \ll (k_1 - k_2)
\end{equation}

The power-of-two component ($k_1 - k_2$) is computed via subtraction and shift operations following leading one detector (LOD). Both operands first undergo LOD processing to determine the exponent values $k_i$, where LOD($X$) = $k_1$ denotes the position of the most significant '1' in $X$.

The fractional division ($x/y$) is implemented using a two-dimensional lookup table (2D-LUT). After normalization, the four most significant bits (MSBs) following the leading '1' of $x$ and $y$ serve as row and column indices, respectively. This 4$\times$4-bit indexing confines the LUT to 256 entries, providing 8-bit precision for the fractional quotient.

The LOD module employs a hierarchical binary search algorithm. For $k$-bit inputs, the $\log_2k$-stage process successively examines 8-bit, 4-bit, and finer segments. At each stage, the presence of '1' in the upper half determines segment selection, as shown in Algorithm~\ref{alg:lod}. This approach reduces logic depth by 58\% compared to sequential detection for 16-bit operands.

The final quotient combines the fractional component from the LUT with the exponent-driven shift, completing the division in three pipelined stages: (1) normalization and LOD (2) fractional division (3) recombination and denormalization.

\begin{figure*}[t] 
\captionsetup{font=small}
\centering
\includegraphics[width=140mm]{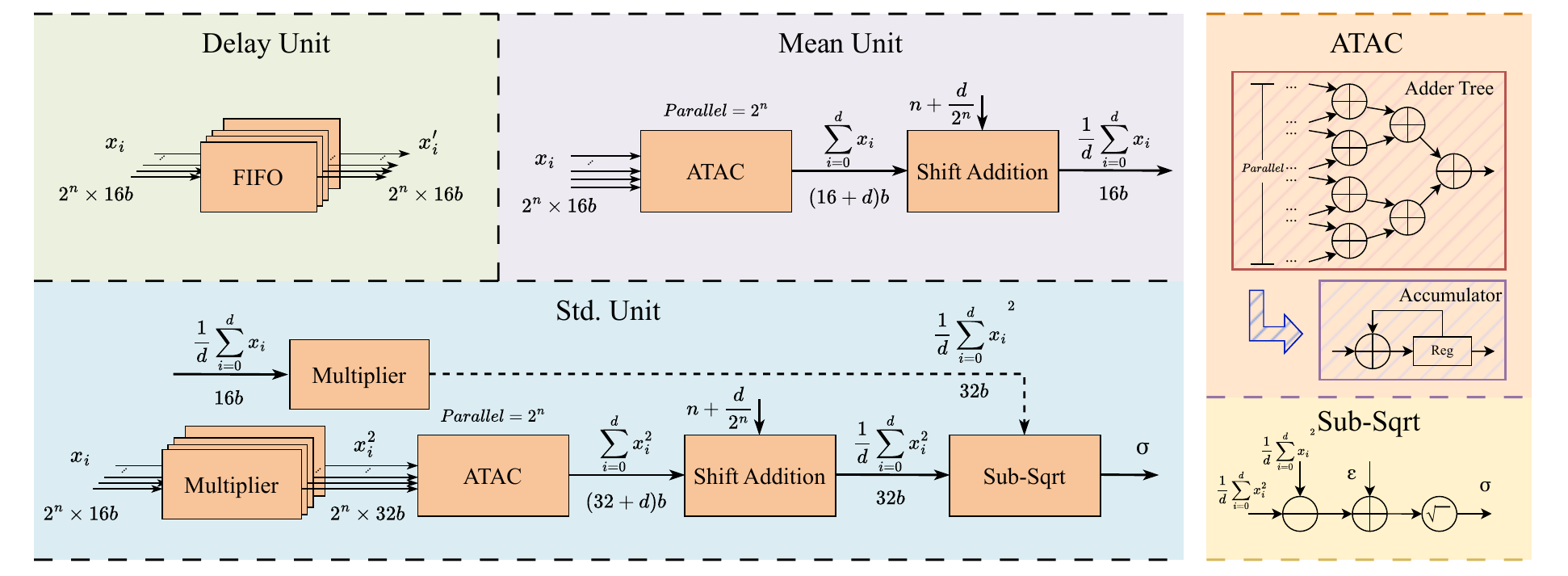}
\caption{LayerNorm Module}
\label{LN}
\vspace{-0.2cm}
\end{figure*}

\subsection{Exponential–Sigmoid Unit}

In the proposed design, the reusable exponential–sigmoid (EXP–$\sigma$) unit supports two distinct operations—base‑$e$ exponentiation and sigmoid activation—under a single unified architecture (Figure~\ref{DIVU_EXP_Sigmoid}(b)). By selecting the control signal $mode$, the same datapath can be repurposed, reducing area and control overhead. 

When $mode=0$, the module exploits a mathematical transformation to convert natural exponentiation into base‑2 form\cite{EXP}, thereby leveraging efficient shift operations:
\begin{equation}
 e^{X}=2^{Y},\quad Y=X \log _{2} e,\quad \log _{2} e \approx 1.0111_2 
\end{equation}
The product \(Y\) is then decomposed into its integer part \(u\) and fractional part \(v\). The term \(2^u\) is calculated through bit‑shifting, while the first eight bits of \(v\) index into a 256‑entry EXP‑LUT to yield eight‑bit precision. Multiplication by the constant \(\log_2e\) is realized using a single addition, one subtraction, and two shift operations.

When $mode=1$, the module computes the sigmoid function by a piecewise linear approximation\cite{Sigmoid}, wherein each segment’s slope is decomposed into a sum of dyadic fractions and implemented via the shared ShiftAddition unit. The required coefficients are retrieved from the $\sigma$‑LUT. The approximation is defined as
\begin{equation}
f(x)=
\begin{cases}
1, & x\ge5\\
0.03125\,x+0.84375, & 2.375\le x <5\\
0.125\,x+0.625, & 1\le x <2.375\\
0.25\,x+0.5, & 0\le x <1\\
1-f(-x), & x<0
\end{cases}
\end{equation}
This method ensures both high throughput and low logic utilization, as the heavy lifting of nonlinear evaluation is offloaded to small, fast memories.

In both modes, all fixed‑constant multiplications—whether by \(\log_2e\) or by segment slopes—are replaced by a dedicated ShiftAddition unit. This unit dynamically selects and combines bit‑shifted operands with minimal arithmetic circuitry, thereby maximizing reuse across modes and simplifying the overall control logic.

\subsection{LayerNorm Module}

In many previous studies, the computational workload of the LayerNorm layer has been delegated to the CPU\cite{LN_CPU}, owing to its capability to handle complex operations such as division and standard‐deviation calculation more natively. However, this approach incurs data‐movement latency, reduces overall system throughput, and exacerbates memory‐bandwidth pressure. Noting that the bulk of the LayerNorm computation consists of vector summations—operations that FPGAs excel at—we have designed a fully on‑chip LayerNorm module to eliminate off‑chip transfers and realize end‑to‑end acceleration.\cite{LN}

The LayerNorm computation requires the mean
\begin{equation}
\mu = \frac{1}{d} \sum_{i=1}^d x_i
\end{equation}
and variance
\begin{equation}
\sigma^2 = \frac{1}{d} \sum_{i=1}^d \bigl(x_i - \mu\bigr)^2
\end{equation}
where \(x_i\) denotes the \(i\)th element of the input vector and \(d\) is its length. By applying the identity
\begin{equation}
\sigma^2 = \frac{1}{d}\sum_{i=1}^d x_i^2 \;-\;\Bigl(\frac{1}{d}\sum_{i=1}^d x_i\Bigr)^2
\end{equation}
we reduce two separate summations—one of \(x_i\) and one of \(x_i^2\)—followed by a subtraction and a square‐root:
\begin{equation}
\sigma = \sqrt{\sigma^2 + \epsilon}
\end{equation}
where \(\epsilon\) is a small constant to ensure numerical stability.

To accelerate these summations on FPGA, we employ a pipelined addition tree (AT) in conjunction with an accumulator (AC), together termed the ATAC structure. With a tree parallelism of 512, each summation of \(d\) elements completes in \(\lceil d/512 \rceil + 9\) clock cycles (Figure~\ref{LN}). The input vector \(X\) is partitioned into blocks matching the ATAC’s parallel width; both the mean and variance paths contain identical ATAC units. Once each ATAC has processed all blocks, its output is held constant. In the mean path, the final accumulated sum is divided by the constant \(d\) via optimized shift‑and‑add operations. In the variance path, the mean is fanned out and squared, while each \(x_i\) is squared and then reduced by the mean square. These two streams feed a subtract–square‑root module that computes the variance, adds \(\epsilon\), and takes the square root to yield \(\sigma\). A delay buffer ensures synchronization so that, upon arrival of the first block at the normalization stage, both \(\mu\) and \(\sigma\) are already valid. Finally, each block of \(X\) is normalized by computing \((x_i - \mu)/\sigma\) and streamed out (cf.\ Figure~\ref{System Architecture}).

\begin{table*}[t]
\centering
\caption{Perplexity (ppl) and accuracy (acc) of RWKV under different quantization schemes and datasets.}
\label{PPL}
\resizebox{\textwidth}{!}{
\begin{tabular}{lccccccccc}
\toprule
\multirow{2}{*}{\textbf{Precision}} 
& \textbf{LAMBADA}              & \textbf{LAMBADA}              & \textbf{HellaSwag}     
& \textbf{Arc-Easy}             & \textbf{Arc-Challenge }       & \textbf{SciQ} 
& \textbf{PIQA}             & \textbf{Winogrande}           & \textbf{Average} \\
& \textbf{ppl $\downarrow$}     & \textbf{acc $\uparrow$}       & \textbf{acc $\uparrow$}     
& \textbf{acc $\uparrow$}       & \textbf{acc $\uparrow$}       & \textbf{acc $\uparrow$} 
& \textbf{acc $\uparrow$}       & \textbf{acc $\uparrow$}       & \textbf{acc $\uparrow$} \\
\midrule
\bfseries FP16      & 7.18      & 58.00       & 42.33       & 47.0        & 23.6       & 82.6   & 72.2 & 54.6 & 54.33\\
\midrule
\bfseries RTN       & 8.40        & 52.29        & 40.33       & 46.1      & 22.3       & 80.4   & 70.4 & 52.4 & 52.03 \\
\bfseries PoT       & 14.4       & 43.14       & 37.14         & 45.5      & 22.2    & 78.8   & 69.6 & 51.6 & 49.43 \\
\bfseries LogQ      & 8.28        & 52.71       & 40.11        & 46.2      & 22.6        & 80.2  & 70.5 & 52.6 & 52.13 \\
\bfseries Proposed  & 7.24        & 56.41          & 41.33       & 46.2     & 22.9        & 81.4  & 70.4 & 53.2 & 53.12 \\
\bottomrule
\end{tabular}
}
\end{table*}

\section{EXPERIMENTS}
\subsection{Experimental Setup}
\paragraph{\textbf{Model and datasets.}} To systematically evaluate the impact of different quantization strategies on RWKV model performance, we follow established evaluation protocols by applying our mixed-precision quantization scheme alongside comparative quantization methods (RTN \cite{uniform_quantization}, LogQ \cite{Log_Quantization}\cite{Log_Quantization_2}, PoT \cite{PoT}) to the RWKV architecture. Model effectiveness is assessed through Perplexity(ppl) on the LAMBADA dataset\cite{dataset_LAMBADA} and zero-shot accuracy(acc) across seven benchmark datasets: LAMBADA, HellaSwag \cite{dataset_Hella}, Arc-Easy \cite{dataset_ARC}, Arc-Challenge \cite{dataset_ARC}, SciQ \cite{dataset_SciQ}, PIQA \cite{dataset_PIQA}, and Winogrande \cite{dataset_WinoGrande}. Experimental results are summarized in Table~\ref{PPL}.

\paragraph{\textbf{Implementation.}}The proposed design was implemented on Xilinx data center accelerator card platform Alveo U50 called HFRWKV and U280 called HFRWKV*, using Vivado 2023.2. Both platforms are based on the 16nm UltraScale+™ architecture and are equipped with 8 GB of HBM2 high-bandwidth memory. The Alveo U50 provides hardware resources of 872K Look-Up Tables (LUTs), 1,743K registers, 5,952 DSP slices, 1,344 36Kb Block RAMs (BRAMs), 640 288Kb UltraRAMs (URAMs), and a rated bandwidth of 201GB/s.The Alveo U280 provides 1,304K LUTs, 2,607K registers, 9,024 DSP slices, 2,016 36Kb BRAMs, 960 288Kb URAMs, and bandwidth of 460GB/s.

\paragraph{\textbf{Baselines.}}To the best of our knowledge, there are currently no existing FPGA acceleration works dedicated to the RWKV architecture family. Although similar Transformer/RNN accelerators\cite{Accelerator0}\cite{Accelerator2} exist, their structural and algorithmic disparities make performance comparisons do not serve as meaningful benchmarks. Therefore, we adopt CPU and GPU platforms as performance baselines. The CPU reference system employs an Intel Core i7‑12650H (10nm process, 10 cores/16 threads) with 16GB of DDR4 memory. Our GPU baselines cover a spectrum from mid‑range to high‑end: the NVIDIA GeForce RTX 2080Ti (12nm, 11GB), RTX 3090 (8nm, 24GB), and the A100 machine‑learning‑optimized HPC card (7nm, 40GB). For both CPU and GPU measurements, we used the official RWKV implementation of the test platform. Throughput is obtained by repeatedly processing single‑token inputs (batch size=1) and averaging the sustained token‑per‑second rate to provide an objective comparison.

\begin{table}[b]
  \centering
  \fontsize{8.2pt}{8pt}\selectfont 
  \renewcommand{\arraystretch}{1.5} 
  \setlength{\tabcolsep}{0pt}      
  \caption{Resource utilization of HFRWKV and HFRWKV*}
  \label{resources}
  \resizebox{90mm}{!}{
  \begin{tabular}{ccccc}
    \toprule
      & \textbf{HFRWKV\_0} & \textbf{HFRWKV\_1} 
      & \textbf{HFRWKV$^*$\_0} & \textbf{HFRWKV$^*$\_1} \\
    \hline
    \bfseries   Platform      & \multicolumn{2}{c}{Alveo U50} 
                              & \multicolumn{2}{c}{Alveo U280} \\
    \bfseries   Support Size  & 169M               & 430M / 1B5 / 3B / 7B 
                              & 169M               & 430M / 1B5 / 3B / 7B \\
    \bfseries   Frequency     & 350MHz             & 350MHz             
                              & 400MHz             & 400MHz            \\
    \toprule
    \bfseries   LUT           & 95718 (11\%)       & 137631 (16\%)      
                              & 126956 (10\%)      & 182372 (14\%)     \\
    \bfseries   FF            & 82719 (5\%)        & 124350 (7\%)       
                              & 102809 (4\%)       & 151158 (6\%)      \\
    \bfseries   DSP           & 641   (11\%)        & 1025 (17\%)        
                              & 1025 (11\%)        & 1537 (17\%)       \\
    \bfseries   BRAM          & 45    (3\%)         & 637   (47\%)        
                              & 45    (2\%)         & 637   (32\%)       \\
    \bfseries   URAM          & 96    (15\%)        & 128   (20\%)        
                              & 192   (20\%)        & 256   (27\%)       \\
    \toprule
  \end{tabular}
  }
\end{table}

\subsection{Ablation Study on Quantization}


We now conduct the ablation study to show the impact of different quantization schemes on model effectiveness. This ablation study was implemented on GPU using the official rwkv pip package and benchmark scripts from the ChatRWKV project \cite{rwkv_pip_package}\cite{chatrwkv_benchmark_script}. As shown in Table~\ref{PPL}, all results are compared against the non-quantized FP16 baseline in different datasets. Notably, the results for RTN, PoT, and LogQ are measured by simulating the precision loss of an equivalent W9A9 quantization. Applying conventional quantization schemes leads to varying degrees of degradation in performance. The PoT quantization results in the most significant average accuracy drop, down to 49.43\%. While the RTN and LogQ schemes perform better, they still fail to maintain high fidelity, with average accuracy of 52.03\% and 52.13\% respectively, and their ppl for LAMBADA dataset deteriorates to over 8.25.

In contrast, our proposed scheme effectively mitigates this performance loss. It achieves an average accuracy of 53.12\%, clearly outperforming all other quantization schemes. More importantly, our method maintains the ppl for LAMBADA dataset of 7.24, which is substantially closer to the FP16 baseline than any other scheme. This demonstrates that our proposed scheme provides a superior trade-off, preserving model accuracy far more effectively than existing techniques.


\subsection{Hardware Results}
\subsubsection{Resource utilization}

\paragraph{\textnormal{To support models of varying sizes while fully exploiting the high‑bandwidth memory, we implemented two system versions across two Alveo platforms.}} HFRWKV\_0(HFRWKV*\_0) is applied to the smallest model, while HFRWKV\_1(HFRWKV*\_1) handles the remaining models. HFRWKV were implemented on the Alveo U50, achieving peak bandwidth utilization of 99.95\% at 350 MHz on-chip frequency. HFRWKV* was deployed on the Alveo U280, attaining 99.64\% bandwidth utilization at 400 MHz. In HFRWKV\_0, the Matrix-Vector Processing Array sets $l$ as the dimension of weight vectors with batch parameter $d=384$, while its LayerNorm Module configures das the weight vector dimension with $tree \; parallelism=256$. For [HFRWKV\_1, HFRWKV*\_0 ,HFRWKV*\_1], their configurations are $d=[512, 768, 1024]$ and $tree \; parallelism=[512, 256, 512]$. All four configurations integrate 128 replicated Unsigned Division Units and Exponential–Sigmoid Units. Such parameter design adapts to varying model sizes and maximizes on-board resource and bandwidth utilization. The corresponding resource utilization metrics are detailed in Table~\ref{resources}.

\begin{figure}[h] 
\captionsetup{font=small}
\centering
\includegraphics[width=85mm]{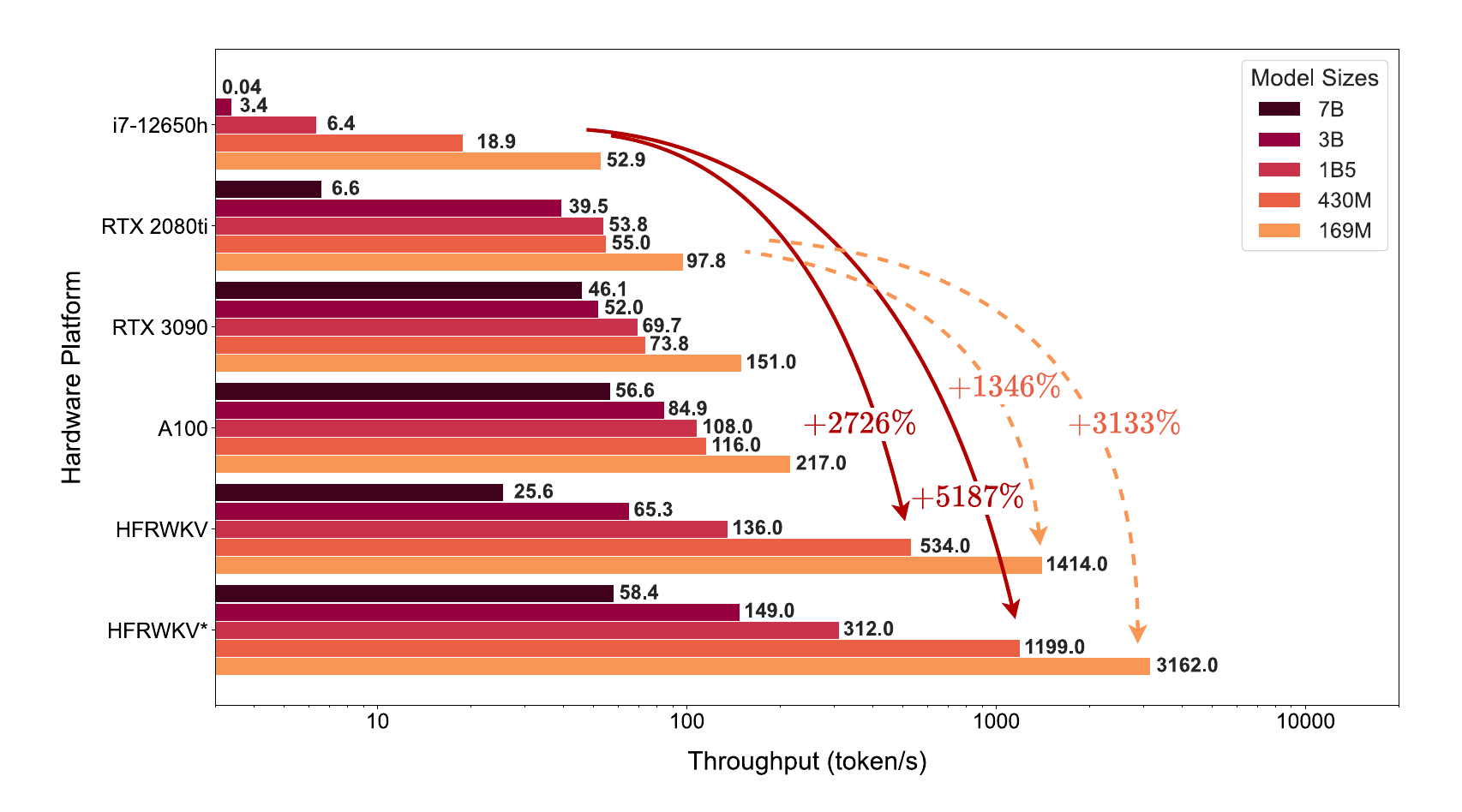}
\caption{Throughput of CPU, GPUs, HFRWKV and HFRWKV*}
\label{Throughput}
\vspace{-0.2cm}
\end{figure}

\begin{figure}[h] 
\captionsetup{font=small}
\centering
\includegraphics[width=85mm]{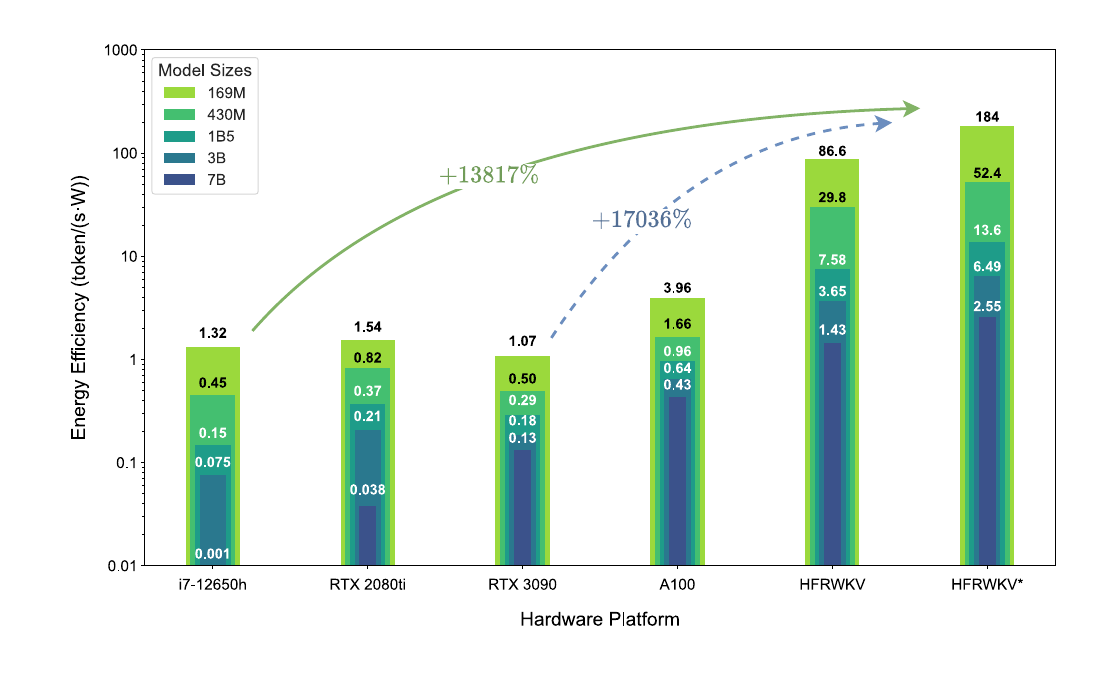}
\caption{Energy Efficient of CPU, GPUs, HFRWKV and HFRWKV*}
\label{Energy_efficient}
\vspace{-0.1cm}
\end{figure}

\subsubsection{Performance comparison}
\paragraph{\textbf{Speedup.}} Throughput measurements for RWKV model inference across platforms are summarized in Fig.~\ref{Throughput}. For the model size of 169M, HFRWKV attains 26.74× the throughput of the CPU baseline and 14.46×, 9.37×, and 6.51× those of the RTX 2080Ti, RTX 3090, and A100, respectively. HFRWKV* further increases these factors to 59.8×, 32.33×, 20.95×, and 14.55×.  

For the largest model size of 7B, HFRWKV achieves only 0.55× the throughput of the RTX3090 and 0.45× that of the A100. In contrast, HFRWKV* slightly surpasses the A100, reaching 1.03× its throughput, and thus outperforms the A100 across all model scales.

		

\paragraph{\textbf{Energy efficiency.}}
 Power measurements derived from Vivado power analysis reports reveal significant energy efficiency advantages. As shown in Fig.~\ref{Energy_efficient}, both HFRWKV and HFRWKV* exhibit substantially higher energy efficiency compared to the CPU and GPU baselines, underscoring the benefits of our FPGA‑based design.

	
			

\section{CONCLUSION}

This paper proposes HFRWKV, a RWKV hardware accelerator with full on-chip computing architecture employing mixed-precision quantization. We propose a novel $\Delta$-PoT quantization for multiplication operations and devise a novel hardware-friendly hybrid-precision quantization strategy. For complex operations including exponentiation and division, we introduce a method featuring reusable architectures combined with lookup tables or piecewise linear approximation. Building upon these innovations, we constructed a high-throughput pipelined architecture integrating a parallel matrix–vector processing array, complex computing units and on-chip LayerNorm module. Through computational reordering and chunked double buffering techniques, our architecture eliminates data transfer bottlenecks and enables complete on-chip computation. When accelerating RWKV‑4, HFRWKV attains 63.48× speedup and 139.17× energy efficiency improvement over a CPU baseline, and 32.33× speedup with 171.36× better energy efficiency compared to GPU implementations.

Although this paper focuses on RWKV‑4, HFRWKV is inherently compatible with the entire RWKV model family. In future work, we will extend our FPGA acceleration framework to support all RWKV variants.

\section{Acknowledgments}

This work was supported by the National Natural Science Foundation of China under Grant 62276278 and GuangDong Basic and Applied Basic Research Foundation under Grant 2022A1515110006 and 2024A1515011259.

\bibliographystyle{unsrt}
\bibliography{sample-base}

@String{Computing = "Computing" }

@String{Computer = "{IEEE} Computer" }

@String{NIPS = "Advances in Neural Information Processing Systems"}

@String{WACV = "Proceedings of the {IEEE/CVF} Winter Conference on Applications of Computer Vision"}

@String{ICML = "International Conference on Machine Learning"}

@String{ICASSP = "IEEE International Conference on Acoustics, Speech and Signal Processing ({ICASSP})"}

@String{APCCAS = "IEEE Asia Pacific Conference on Circuits and Systems ({APCCAS})"}

@String{FPL = "International Conference on Field-Programmable Logic and Applications ({FPL})"}

@String{HIPC = "IEEE International Conference on High Performance Computing, Data, and Analytics ({HiPC})"}

@String{IAIT = "International Conference on Advances in Information Technology"}

@String{arXiv = "{arXiv} preprint"}

@String{CoRR = "{CoRR}"}

@String{IEEECircSys = "{IEE} Proceedings-Circuits, Devices and Systems"}

@String{AppliedSci = "Applied Sciences"}

@String{Curran = "Curran Associates, Inc."}

@String{PMLR = "{PMLR}"}

@String{IEEE = "{IEEE}"}

@String{IET = "{IET}"}

@inproceedings{transformer,
  author    = {Vaswani, Ashish and Shazeer, Noam and Parmar, Niki and Uszkoreit, Jakob and Jones, Llion and Gomez, Aidan N and Kaiser, {\L}ukasz and Polosukhin, Illia},
  booktitle = NIPS,
  editor    = {I. Guyon and U. Von Luxburg and S. Bengio and H. Wallach and R. Fergus and S. Vishwanathan and R. Garnett},
  pages     = {},
  publisher = Curran,
  title     = {Attention is All you Need},
  url       = {https://proceedings.neurips.cc/paper_files/paper/2017/file/3f5ee243547dee91fbd053c1c4a845aa-Paper.pdf},
  volume    = {30},
  year      = {2017}
}

@article{transformer1,
  title     = {{Opt: Open pre-trained transformer language models}},
  author    = {Zhang, Susan and Roller, Stephen and Goyal, Naman and Artetxe, Mikel and Chen, Moya and Chen, Shuohui and Dewan, Christopher and Diab, Mona and Li, Xian and Lin, Xi Victoria and others},
  journal   = arXiv # " arXiv:2205.01068",
  year      = {2022}
}

@misc{LLM_infuluence,
  title        = {A Survey on Large Language Models with some Insights on their Capabilities and Limitations},
  author       = {Andrea Matarazzo and Riccardo Torlone},
  year         = {2025},
  eprint       = {2501.04040},
  archivePrefix= {arXiv},
  primaryClass = {cs.CL},
  url          = {https://arxiv.org/abs/2501.04040},
}

@InProceedings{linear_attention,
  author    = {Shen, Zhuoran and Zhang, Mingyuan and Zhao, Haiyu and Yi, Shuai and Li, Hongsheng},
  title     = {Efficient Attention: Attention With Linear Complexities},
  booktitle = WACV,
  month     = {January},
  year      = {2021},
  pages     = {3531-3539}
}

@inproceedings{linear_attention1,
  title        = {Transformers are rnns: Fast autoregressive transformers with linear attention},
  author       = {Katharopoulos, Angelos and Vyas, Apoorv and Pappas, Nikolaos and Fleuret, Fran{\c{c}}ois},
  booktitle    = ICML,
  pages        = {5156--5165},
  year         = {2020},
  organization = PMLR
}

@article{rwkv-4,
  title     = {{Rwkv: Reinventing rnns for the transformer era}},
  author    = {Peng, Bo and Alcaide, Eric and Anthony, Quentin and Albalak, Alon and Arcadinho, Samuel and Biderman, Stella and Cao, Huanqi and Cheng, Xin and Chung, Michael and Grella, Matteo and others},
  journal   = arXiv # " arXiv:2305.13048",
  year      = {2023}
}

@misc{layernorm,
  title        = {Layer Normalization},
  author       = {Jimmy Lei Ba and Jamie Ryan Kiros and Geoffrey E. Hinton},
  year         = {2016},
  eprint       = {1607.06450},
  archivePrefix= {arXiv},
  primaryClass = {stat.ML},
  url          = {https://arxiv.org/abs/1607.06450},
}

@article{uniform_quantization,
  author    = {Benoit Jacob and Skirmantas Kligys and Bo Chen and Menglong Zhu and Matthew Tang and Andrew G. Howard and Hartwig Adam and Dmitry Kalenichenko},
  title     = {Quantization and Training of Neural Networks for Efficient Integer-Arithmetic-Only Inference},
  journal   = CoRR,
  volume    = {abs/1712.05877},
  year      = {2017},
  url       = {http://arxiv.org/abs/1712.05877},
  eprinttype= {arXiv},
  eprint    = {1712.05877}
}

@Article{non_uniform_quantization,
  author        = {Seo, Sanghyun and Kim, Juntae},
  title         = {Efficient Weights Quantization of Convolutional Neural Networks Using Kernel Density Estimation based Non-uniform Quantizer},
  journal       = AppliedSci,
  volume        = {9},
  year          = {2019},
  number        = {12},
  article-number= {2559},
  url           = {https://www.mdpi.com/2076-3417/9/12/2559},
  doi           = {10.3390/app9122559}
}

@article{non_uniform_quantization_2,
  author    = {Dongqing Zhang and Jiaolong Yang and Dongqiangzi Ye and Gang Hua},
  title     = {{LQ-Nets: Learned Quantization for Highly Accurate and Compact Deep Neural Networks}},
  journal   = CoRR,
  volume    = {abs/1807.10029},
  year      = {2018},
  url       = {http://arxiv.org/abs/1807.10029},
  eprinttype= {arXiv},
  eprint    = {1807.10029}
}

@article{non_uniform_quantization_3,
  author    = {Antonio Polino and Razvan Pascanu and Dan Alistarh},
  title     = {Model compression via distillation and quantization},
  journal   = CoRR,
  volume    = {abs/1802.05668},
  year      = {2018},
  url       = {http://arxiv.org/abs/1802.05668},
  eprinttype= {arXiv},
  eprint    = {1802.05668}
}

@inproceedings{Log_Quantization,
  author    = {Cai, Jingyong and Takemoto, Masashi and Nakajo, Hironori},
  title     = {A Deep Look into Logarithmic Quantization of Model Parameters in Neural Networks},
  year      = {2018},
  publisher = "Association for Computing Machinery",
  booktitle = IAIT,
  series    = {IAIT '18}
}

@INPROCEEDINGS{Log_Quantization_2,
  author    = {Lee, Edward H. and Miyashita, Daisuke and Chai, Elaina and Murmann, Boris and Wong, S. Simon},
  booktitle = ICASSP,
  title     = {{LogNet: Energy-efficient neural networks using logarithmic computation}},
  year      = {2017},
  pages     = {5900-5904},
  doi       = {10.1109/ICASSP.2017.7953288}
}

@article{PoT,
  author    = {Aojun Zhou and Anbang Yao and Yiwen Guo and Lin Xu and Yurong Chen},
  title     = {Incremental Network Quantization: Towards Lossless CNNs with Low-Precision Weights},
  journal   = CoRR,
  volume    = {abs/1702.03044},
  year      = {2017},
  url       = {http://arxiv.org/abs/1702.03044},
  eprinttype= {arXiv},
  eprint    = {1702.03044}
}

@article{MIX_Q_inHardware_difficulty,
  author    = {Yang Lin and Tianyu Zhang and Peiqin Sun and Zheng Li and Shuchang Zhou},
  title     = {{FQ-ViT: Fully Quantized Vision Transformer without Retraining}},
  journal   = CoRR,
  volume    = {abs/2111.13824},
  year      = {2021},
  url       = {https://arxiv.org/abs/2111.13824},
  eprinttype= {arXiv},
  eprint    = {2111.13824}
}

@article{APoT,
  author    = {Yuhang Li and Xin Dong and Wei Wang},
  title     = {Additive Powers-of-Two Quantization: {A} Non-uniform Discretization for Neural Networks},
  journal   = CoRR,
  volume    = {abs/1909.13144},
  year      = {2019},
  url       = {http://arxiv.org/abs/1909.13144},
  eprinttype= {arXiv},
  eprint    = {1909.13144}
}

@article{nhot_PoT,
  author    = {Yuiko Sakuma and Hiroshi Sumihiro and Jun Nishikawa and Toshiki Nakamura and Ryoji Ikegaya},
  title     = {n-hot: Efficient bit-level sparsity for powers-of-two neural network quantization},
  journal   = CoRR,
  volume    = {abs/2103.11704},
  year      = {2021},
  url       = {https://arxiv.org/abs/2103.11704},
  eprinttype= {arXiv},
  eprint    = {2103.11704}
}

@ARTICLE{Systolic_Array,
  author   = {Kung},
  journal  = Computer,
  title    = {Why systolic architectures?},
  year     = {1982},
  volume   = {15},
  number   = {1},
  pages    = {37-46},
  doi      = {10.1109/MC.1982.1653825}
}

@INPROCEEDINGS{EXP,
  author    = {Wang, Meiqi and Lu, Siyuan and Zhu, Danyang and Lin, Jun and Wang, Zhongfeng},
  booktitle = APCCAS,
  title     = {A High-Speed and Low-Complexity Architecture for Softmax Function in Deep Learning},
  year      = {2018},
  pages     = {223-226},
  doi       = {10.1109/APCCAS.2018.8605654}
}

@article{Sigmoid,
  title     = {Piecewise linear approximation applied to nonlinear function of a neural network},
  author    = {Amin, Hesham and Curtis, K Memy and Hayes-Gill, Barrie R},
  journal   = IEEECircSys,
  volume    = {144},
  number    = {6},
  pages     = {313--317},
  year      = {1997},
  publisher = IET
}

@INPROCEEDINGS{LN_CPU,
  author    = {Li, Zhengang and Sun, Mengshu and Lu, Alec and Ma, Haoyu and Yuan, Geng and Xie, Yanyue and Tang, Hao and Li, Yanyu and Leeser, Miriam and Wang, Zhangyang and Lin, Xue and Fang, Zhenman},
  booktitle = FPL,
  title     = {{Auto-ViT-Acc: An FPGA-Aware Automatic Acceleration Framework for Vision Transformer with Mixed-Scheme Quantization}},
  year      = {2022},
  pages     = {109-116},
  doi       = {10.1109/FPL57034.2022.00027}
}

@inproceedings{LN,
  title        = {{Me-vit: A single-load memory-efficient fpga accelerator for vision transformers}},
  author       = {Marino, Kyle and Zhang, Pengmiao and Prasanna, Viktor K},
  booktitle    = HIPC,
  pages        = {213--223},
  year         = {2023},
  organization = IEEE
}

@article{Accelerator0,
  author    = {Bingbing Li and Santosh Pandey and Haowen Fang and Yanjun Lyv and Ji Li and Jieyang Chen and Mimi Xie and Lipeng Wan and Hang Liu and Caiwen Ding},
  title     = {{FTRANS:} Energy-Efficient Acceleration of Transformers using {FPGA}},
  journal   = CoRR,
  volume    = {abs/2007.08563},
  year      = {2020},
  url       = {https://arxiv.org/abs/2007.08563},
  eprinttype= {arXiv},
  eprint    = {2007.08563}
}

@INPROCEEDINGS{Accelerator2,
  author    = {Han, Yuntao and Liu, Qiang},
  booktitle = FPL,
  title     = {{HPTA: A High Performance Transformer Accelerator Based on FPGA}},
  year      = {2023},
  pages     = {27-33},
  doi       = {10.1109/FPL60245.2023.00012}
}

@article{dataset_LAMBADA,
  author    = {Denis Paperno and Germ{\'{a}}n Kruszewski and Angeliki Lazaridou and Quan Ngoc Pham and Raffaella Bernardi and Sandro Pezzelle and Marco Baroni and Gemma Boleda and Raquel Fern{\'{a}}ndez},
  title     = {The {LAMBADA} dataset: Word prediction requiring a broad discourse context},
  journal   = CoRR,
  volume    = {abs/1606.06031},
  year      = {2016},
  url       = {http://arxiv.org/abs/1606.06031},
  eprinttype= {arXiv},
  eprint    = {1606.06031}
}

@article{dataset_Hella,
  author    = {Rowan Zellers and Ari Holtzman and Yonatan Bisk and Ali Farhadi and Yejin Choi},
  title     = {{HellaSwag: Can a Machine Really Finish Your Sentence?}},
  journal   = CoRR,
  volume    = {abs/1905.07830},
  year      = {2019},
  url       = {http://arxiv.org/abs/1905.07830},
  eprinttype= {arXiv},
  eprint    = {1905.07830}
}

@article{dataset_ARC,
  author    = {Peter Clark and Isaac Cowhey and Oren Etzioni and Tushar Khot and Ashish Sabharwal and Carissa Schoenick and Oyvind Tafjord},
  title     = {Think you have Solved Question Answering? Try ARC, the {AI2} Reasoning Challenge},
  journal   = CoRR,
  volume    = {abs/1803.05457},
  year      = {2018},
  url       = {http://arxiv.org/abs/1803.05457},
  eprinttype= {arXiv},
  eprint    = {1803.05457}
}

@article{dataset_SciQ,
  author    = {Johannes Welbl and Nelson F. Liu and Matt Gardner},
  title     = {Crowdsourcing Multiple Choice Science Questions},
  journal   = CoRR,
  volume    = {abs/1707.06209},
  year      = {2017},
  url       = {http://arxiv.org/abs/1707.06209},
  eprinttype= {arXiv},
  eprint    = {1707.06209}
}

@article{dataset_PIQA,
  author    = {Yonatan Bisk and Rowan Zellers and Ronan Le Bras and Jianfeng Gao and Yejin Choi},
  title     = {{PIQA:} Reasoning about Physical Commonsense in Natural Language},
  journal   = CoRR,
  volume    = {abs/1911.11641},
  year      = {2019},
  url       = {http://arxiv.org/abs/1911.11641},
  eprinttype= {arXiv},
  eprint    = {1911.11641}
}

@article{dataset_WinoGrande,
  author    = {Sakaguchi, Keisuke and Bras, Ronan Le and Bhagavatula, Chandra and Choi, Yejin},
  title     = {WinoGrande: an adversarial winograd schema challenge at scale},
  year      = {2021},
  publisher = "Association for Computing Machinery",
  journal   = "Commun. ACM",
  month     = aug,
  pages     = {99--106}
}

@misc{rwkv_pip_package,
  author       = {BlinkDL},
  title        = {The official rwkv pip package},
  howpublished = {\url{https://pypi.org/project/rwkv/}}
}

@misc{chatrwkv_benchmark_script,
  author       = {BlinkDL},
  title        = {The ChatRWKV benchmark script},
  howpublished = {\url{https://github.com/BlinkDL/ChatRWKV/blob/main/v2/benchmark.py}}
}

\end{document}